\documentclass[pre,twocolumn,aps,groupedaddress]{revtex4}
\usepackage[T1]{fontenc}
\usepackage[latin9]{inputenc}
\usepackage{array}
\usepackage{booktabs}
\usepackage{mathrsfs}
\usepackage{mathbbol}
\usepackage{multirow}
\usepackage{amsmath}
\usepackage{amsthm}
\usepackage{amssymb}
\usepackage{stmaryrd}
\usepackage{graphicx}
\usepackage{latexsym}
\usepackage{textcomp}
\usepackage{mathtools}
\usepackage{cancel}
\usepackage{xcolor}
\usepackage{soul}
\usepackage[allcolors=blue,colorlinks=true]{hyperref}
\usepackage{lineno}
\usepackage{stackrel}
\usepackage[toc,page,titletoc]{appendix}
\usepackage{bm}
\definecolor{mycolor1}{rgb}{0.1, 0.6, 0.6}

\begin{document}

\title{Shear jamming transition in alternating shear rotation for frictional and frictionless suspensions}

\author{Pappu Acharya}
\email{pappu.acharya@univ-grenoble-alpes.fr}
\affiliation{Universit\'e Grenoble Alpes, CNRS, LIPhy, 38000 Grenoble, France}

\affiliation{Computational Chemistry, Lund University, Lund SE-221 00, Sweden}

\author{Martin Trulsson}
\email{martin.trulsson@compchem.lu.se}

\affiliation{Computational Chemistry, Lund University, Lund SE-221 00, Sweden}

\begin{abstract}
\textcolor{black}{Alternating shear rotations in dense suspensions have recently shown the ability to reduce both viscosity and dissipation per strain (at a fixed global shear rate).
Here, we study alternating shear rotation, with extensive numerical simulations, at various angles and up to their corresponding jamming points. For increasing shear rotation angles, we find that the jamming point is continuously shifted to higher packing fractions for frictional particles, while it remains constant for frictionless particles. As a consequence, the alternating shear rotation is unable to reduce the dissipation per strain for suspensions composed of frictionless particles.   
We detail the individual contributions, hydrodynamic or contact, to the shear stress, being uncharted for this protocol. As the angle of rotation increases, the average contact stress decreases. However, we find that the hydrodynamics shows the opposite trend, instead increasing with increasing angle. Hence, hydrodynamic stress will dominate up to much higher packing fractions as the angle of rotation increases. In addition, we report how the microstructure varies and establish a one-to-one mapping between the contact number and its contribution to the total stress for both frictionless and frictional particles.} 
\end{abstract}

\pacs{}
\keywords{}

\maketitle

{\it Introduction:}
Suspensions are abundant in nature (\emph{e.g.,} magma), industry (\emph{e.g.,} pill formation), as well as in the human body (\emph{e.g.,} blood). These complex fluids often exhibit non-Newtonian behavior, where the flow resistance is nonlinear as a function of the flow rate, where shear-thinning is characterized by a sub-linear relationship and shear-thickening to a super-linear relationship. 
Non-Newtonian or not, upon increasing the solid content of a suspension, \emph{i.e.,} increasing the solid fraction, $\phi$, the system attains rigidity via a jamming transition when $\phi\to\phi_c$, where the $\phi_c$ is the packing fraction at (shear-)jamming. In general, suspensions show fragility \cite{seto19fragile} where although they are shear jammed in a particular flow direction, a change in the shear flow direction \emph{i.e.,} applying a shear rotation may allow the system to flow \cite{seto19fragile}. Studying shear rotation has emerged as a promising research area for the last few years as it shows significant interesting physics such as history-dependent transverse stress \cite{blanc23fragile} or reducing energy cost \cite{acharya2023optimum} while transporting. Further, the shear rotation potentially challenges \cite{acharya2023optimum} the idea that random organization \cite{ness2018shaken} is the key mechanism which reduces dissipation. In addition to reducing flow resistance, a repeated sudden change of flow and vorticity directions also shifts the jamming point towards larger solid fractions and hits the maximum for a rotation of $180^o$, corresponding to a non-propagating oscillatory shear protocol.
Although the jamming transition \cite{o2003jamming, liu1998jamming} has been extensively studied for various kinds of non-Brownian systems such as granular materials \cite{majmudar2007jamming, zhang2005jamming}, emulsions \cite{zhang2005jamming}, and foams \cite{katgert2013jamming, lespiat2011jamming} \emph{etc.}, the area is largely unexplored for alternating shear rotation, aka tacking \cite{acharya2023optimum}. To fill this knowledge gap, we here study how the jamming transition and individual contributions, contact and short-range hydrodynamics, to the shear stress are altered by the angle in the alternating shear rotation for frictionless and frictional particles.
In this work, we show that the relative viscosity as a function of the packing fraction at a fixed angle of shear rotation follows the Krieger-Dougherty relationships \cite{krieger1959mechanism}. With increasing angle and for frictional grains, we find that the jamming transition shifted to higher packing fractions in a continuous way, while the prefactor of the Krieger-Dougherty relationships decreases. For frictionless grains (and within our precision), the jamming packing fraction remains constant as a function of tacking angle, while the Krieger-Dougherty's prefactor decreases. We also demonstrate that while contact contribution to the total stress decreases with the tacking angle, the hydrodynamic contribution increases, therefore eventually they crosses at some point in the volume fraction axis. We show that this crossing point increases with the angle of shear rotation. In addition, we report how the microstructure varies with tacking angle and packing fraction in the alternating shear rotation.
In the end, we show that the contact contribution of the shear stress versus the number of contacts per particle,
across all packing fractions and angles, can be unified into two master curves: one for frictionless grains and the other for frictional grains, once a geometric factor is taken into account.  

\begin{figure*}[t]
\centering
\includegraphics[width=1.00\linewidth]{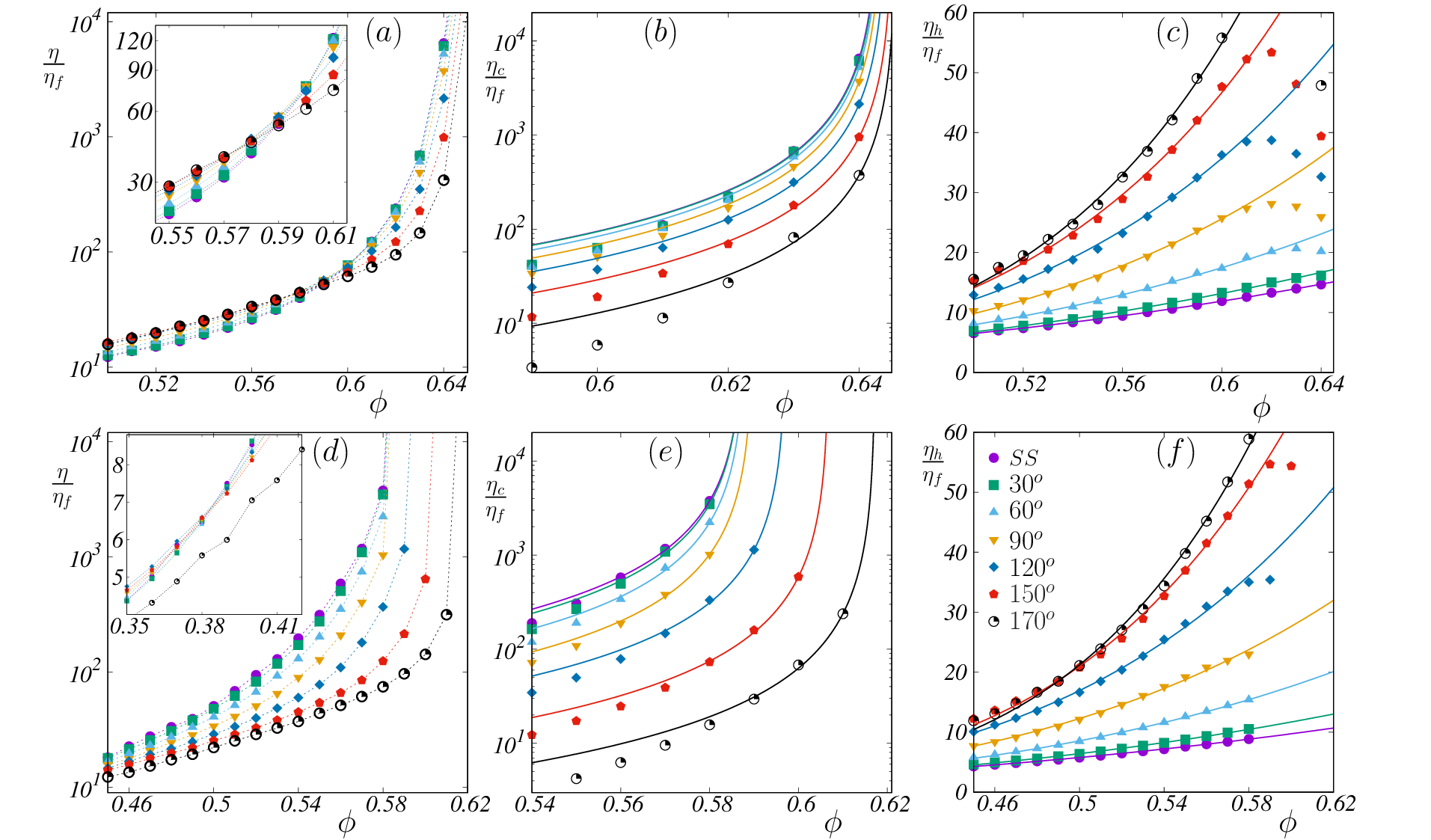}
\caption{(a,d) Relative viscosity along with packing fractions for different shear rotation angle for (a) frictionless and (d) frictional particles. (Inset) Zoomed in version of the figure to show the change in relative viscosity trend from increasing to decreasing with tacking angles. (b,e) Shear jamming, contact contribution on relative viscosity as function of packing fraction for (b) frictionless and (e) frictional particles. The curves are fitted with Krieger-Dougherty relationship. (c,f) Hydrodynamic contribution on relative viscosity as function of packing fraction for (c) frictionless and (f) frictional particles. The branches below jamming are all well-fitted with power law $\alpha \,(\phi/\phi_m)^\beta$, where $\phi_m$ is a reference packing fraction of the fit, \emph{i.e.,} 0.50.}
\label{fig_jamming}
\end{figure*}
\begin{figure}[t!]
\centering

\includegraphics[width=1.0\linewidth]{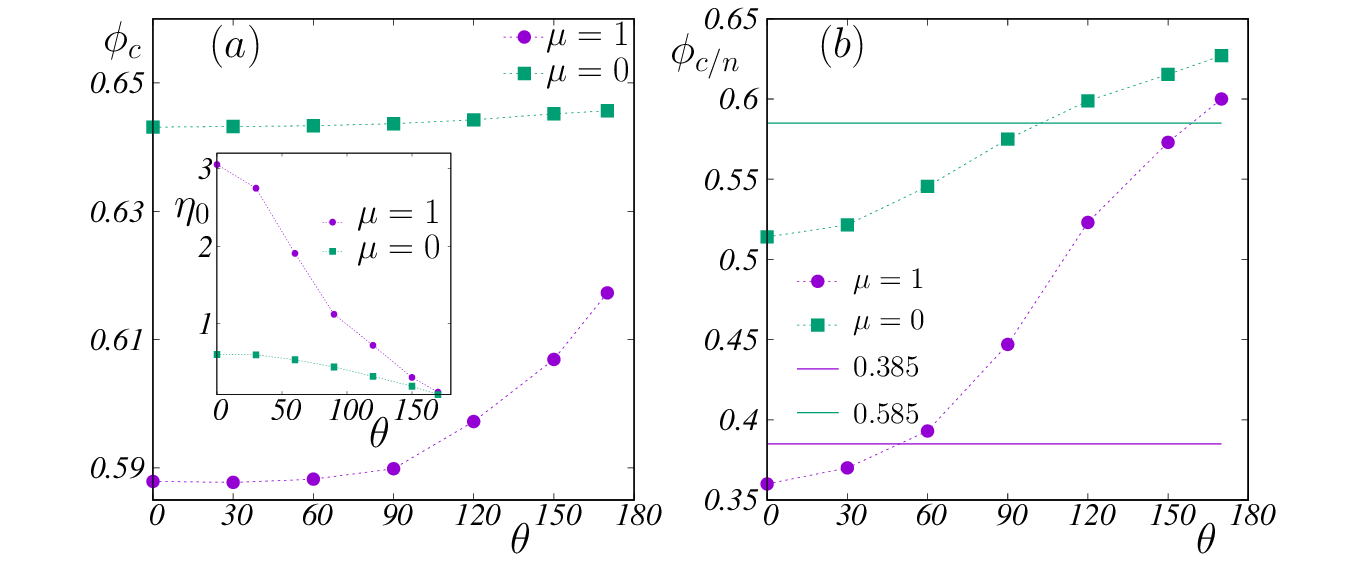}
\caption{(a) Change in jamming point with alternating shear rotation angle for both frictionless and frictional particles by fitting the simulation data with Krieger-Dougherty relationship. While for frictional particle a visible increase is noticed for frictionless the system barely changes its jamming point. (Inset) The prefactor in  Krieger-Dougherty relationship decreases for both frictionless and frictional indicating that the contact contribution decreases in both the cases when a shear rotation is applied.
(b) Crossing packing fraction of hydrodynamics and contact contribution for frictionless and frictional particles. Solid lines corresponds to the crossing point at which the rheology changes behavior as function of angle.   
}
\label{parameter}
\end{figure}
{\it Alternating shear rotation:}
\label{sec_shear_rotation}
A shear rotation, as the name suggests, is a rotation of the flow direction about the gradient axis of a sheared fluid or suspension. As a suspension is sheared, it develops an anisotropic force and contact network \cite{cates1998jamming}, where collisions are more frequent along the compression axis, that increases its resistance to flow. When the flow direction is reversed, this network collapses, resulting in a dip in the viscosity-strain curve. Continuing shearing after a reversal causes the contacts to rebuild, eventually reaching a steady shear state in the new direction. However, frequent changes in direction, particularly reversing after short strain amplitudes, can prevent reaching this steady shear. In a shear rotation, unlike for simple reversals, the suspension alters its flow direction by an angle $\theta$, composing a larger family of unsteady shear protocols. Alternating shear rotations are protocols where shear rotations are repeated after certain strains.  The steady shear and shear reversal can be seen as two extreme cases of shear rotations, with $\theta$ being 0 and 180 degrees, respectively. This protocol is similar to the one recently proposed by Blanc \emph{et al.}\cite{blanc23fragile}, but differs on two points: First, our strain is much smaller than in Ref.~\cite{blanc23fragile}, and, hence, our viscosity never relaxes to its steady-shear value. Secondly and related to the first point, this is for repeated shear-rotations.
Upon rotations, the flow and vorticity directions change, and stresses need to be expressed in these new frames (see supplemental materials for details \cite{supplemental}).
 In our simulations (see supplemental materials for details \cite{supplemental}), unless stated otherwise, we subject each direction to a $1\%$ strain before changing direction and where various shear angles is achieved by shearing in two (perpendicular) directions simultaneously. We perform dynamic particle simulations solving trajectories using LAMMPS \cite{plimpton1995fast} under applied shear, with contact forces (including friction) and hydrodynamics included, following previous studies \cite{ness2018shaken, acharya2023optimum, ness2023simulating}. For each angle of shear rotation, we collect data over 36 accumulated strains, after removing the first four accumulated strains using the same protocols.

\begin{figure*}[t!]
\centering
\includegraphics[width=1.0\linewidth]{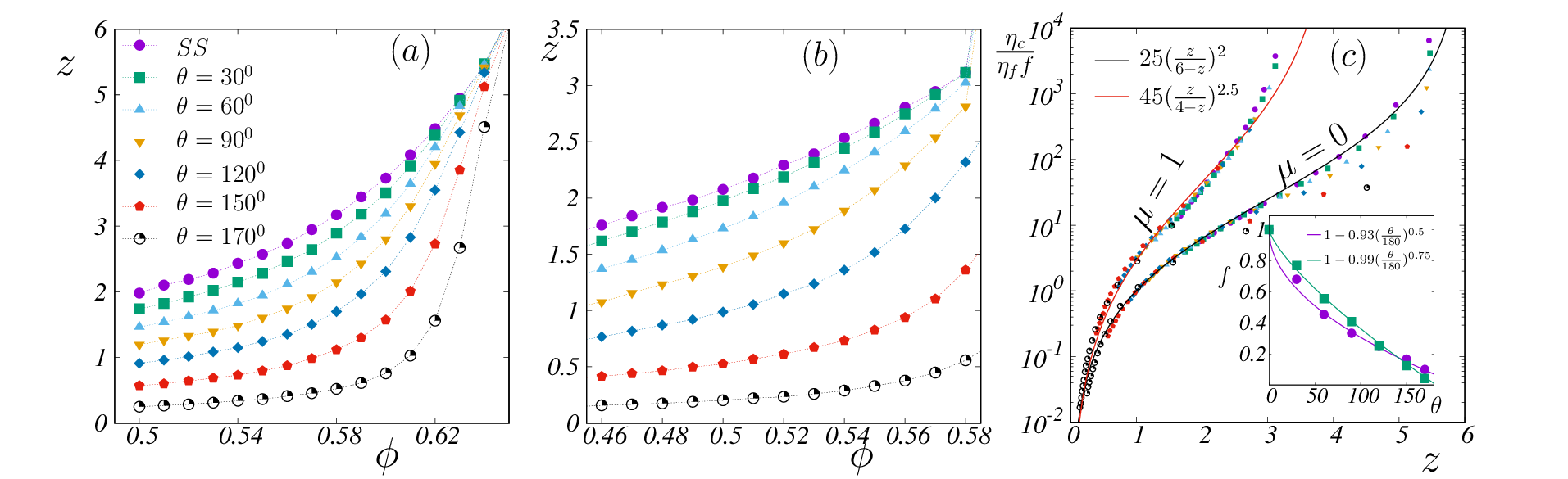}
\caption{(a) The number of average contacts as a function of packing fractions for various shear rotation angles for (a) $\mu=0$ and (b) $\mu=1$. In (a), the curves appear to approach the same jamming point. In contrast, the curves in (b) do not show a tendency to reach the same jamming point. (c) Scaled contact contribution to the viscosity with contact number showing two different branches for frictional and frictionless particles (see Fig. \ref{fig_zvseta} in supplemental material \cite{supplemental} for unscaled contact contributions). (Inset) Factor by which the contact contribution is scaled.}
\label{fig_nn}
\end{figure*}

{\it The jamming transition:}
We now conducted an analysis on the shear jamming transition for non-Brownian suspensions subjected to alternating shear rotations. 
In Fig.~\ref{fig_jamming}(a) and (d), we plot the viscosity with a wide range of packing fractions for different alternating shear rotation angles for frictionless and frictional particles, respectively.
For each fixed angle, we see the viscosity as a function of the packing fraction increasing, as has been observed before in the steady-state conditions.
For the frictionless case, we see that regardless of the angle the viscosity diverges roughly at the same packing fraction of 0.64. This value is the same as found for the isotropic jamming transition, and is identified as the Random Close Packing (RCP), typically around 0.64 in three dimensions \cite{OHern2002}. For the frictional case, the steady-state shear jamming is considerably smaller than in the corresponding frictionless case, 0.59 compared to 0.64 (RCP), due to additional mechanically stabilizing tangential forces \cite{Silbert2010}. As one increases the alternating shear rotation angle the shear-jamming packing fraction shifts to higher values and approaches 0.62, below RCP.
Surprisingly, for frictionless grains, increasing the shear rotation angle at fixed packing fractions alters the viscosity in opposite effects depending on the packing fraction. Below a packing fraction of around 0.585, the viscosity is found to increase with increasing angle, while above it decreases. 
This results in a clear crossing point of the curves at various angles as seen in the inset of Fig.~\ref{fig_jamming}(a). 
We find similar findings (expect for the highest angle) for the frictional case, but where the crossing now is shifted to much smaller packing fractions around 0.385 as can be seen in the inset of Fig.~\ref{fig_jamming}(d).
We now examine the individual contributions of contact forces and hydrodynamics to viscosity, $\eta_c$ and $\eta_h$, with increasing volume fraction at various angle of shear rotations; see Figs.~\ref{fig_jamming}(b, c) for the frictionless case and Figs.~\ref{fig_jamming}(e, f) for the frictional. As can see from the different scales, logarithmic on contact versus linear on hydrodynamics, the contact forces are always orders of greater than the hydrodynamics ones close to its corresponding shear-jamming transition regardless of the angle. However, at lower packing fractions and larger angles these terms starts to be on the same order. We denote the packing fraction where the two contributions, contact and hydrodynamic, have the same magnitude by $\phi_{c/h}$. Interestingly, the two contributions react in oppositely ways upon increasing shear rotation angle.
The contact contribution always decrease upon increasing angle, while the hydrodynamics increase. This explains why we can have the total viscosity increasing at low packing fractions and decreasing at high ones for an increasing angle, as it depends on which contribution is dominate and most responsive. Furthermore, the hydrodynamics do not show any divergence approaching shear-jamming. Instead it shows a non-monotonic behavior as function of packing fraction close the shear-jamming and at the highest angles. 
The contact part of the viscosity can be fitted by using a Krieger-Dougherty relationship, as $\eta_c/\eta_f=\eta_0 (\phi_c-\phi)^{-2.5\phi_c}$, where $\eta_0$ and $\phi_c$ all depends on the rotation angle $\theta$, and are shown in Fig.~\ref{fig_jamming} (b, e) as lines. 
The hydrodynamic contribution is fitted as $\eta_h/\eta_f \sim \alpha (\phi/\phi_m)^\beta$, where $\phi_m$ is a reference packing fraction and we use 0.5 for both frictional and frictionless case. It is important to notice that the hydrodynamic contribution does not diverge as we regularized the minimum distance between any pair of particles, to simulate a particle scale roughness. This clarifies why we use $\phi$ rather than inverse powers of $(\phi_c-\phi)$ in our fits of the hydrodynamics.
The parameters and how they vary with shear rotation angle is illustrated in Fig.~\ref{parameter} and Fig.~\ref{fig_h_para} in the SI \cite{supplemental}. From Fig~\ref{parameter}(a) we see that shear-jamming packing fraction do not show any detectable changes for the frictionless case as the angle is increased, while for the frictional case we see a clear increase when the angle is above $90^\circ$ and is continuously increasing towards its maximum at $\theta\to180^o$, corresponding to alternating shear reversal (or oscillatory shear). The Krieger-Dougherty prefactor $\eta_0$ is seen to decreases with the shear rotation angles for both cases, frictional and frictionless, as seen in the inset of Fig.~\ref{parameter}(a). The fact that $\phi_c$ increases (or stays constant) and $\eta_0$ decreases with increasing cruising angle show that viscosity is lowered with increasing angle for a given fixed packing fraction. It's worth noting that similar trends, increasing shear-jamming packing fraction for frictional grains and a decreased viscosity for both frictional and frictionless grains,  have been observed for oscillatory rheology but as a function of shear amplitude. Here we find similar trends as a function of shear rotation angle. In contrast to pure oscillatory rheology, the tacking protocol consists of a non-zero propagation axis (expect for $\theta=180^\circ$), and particles will change neighbors at one point, hence, excluding the possibility of having reversibility. The reversible-to-irreversible transition (RIT)\cite{pine2005chaos, ge2022rheology} will, hence, ill-defined in case of shear rotation and is only releveant at the singular point of $\theta=180^\circ$.
The hydrodynamic increases, with increasing angle, can be seen from Fig.~\ref{fig_jamming}(c) and (f) but further supported by the fitted parameters seen in Fig.~\ref{fig_h_para}\cite{supplemental}. Next we look at $\phi_{c/h}$ as function of $\theta$, in Fig.~\ref{parameter}(b). 
A compelling observation emerges from our analysis: a tug-of-war scenario unfolds as the contributions from contact forces and lubrication contend with each other. At a certain packing fraction, they intersect, where the contact contribution equals that of lubrication (see Fig. \ref{fig_cross_SI} in SI \cite{supplemental}). Noteworthy is the revelation that the crossing point shifts along the packing fraction axis with the
angle of shear rotation, as evident in Fig.~\ref{parameter}(b). This suggests that as the angle of shear rotation increases, a
higher packing fraction is required for the contact contribution to surpass hydrodynamics. In summary, our
findings emphasize the complex interplay between contact forces and hydrodynamics, shedding light on how
variations in shear rotation angle impact their relative contributions and highlighting the need for a deeper understanding of these dynamics in dense suspensions. 
Surprisingly, the net effect of these opposing forces lead to a almost fixed crossing point in the packing space for the total viscosity (see the insets of Fig.~\ref{fig_jamming}(a) and (d)). 
To evident this a bit better, we also plot the packing fraction at which the total viscosity crosses the steady-state, $\theta=0^\circ$, curve, marked as solid lines in Fig.~\ref{parameter}(b).
Increasing the angle does not only lead to lower viscosity but can also lead to possibility of reduction in the dissipation (per propagation strain), \emph{i.e.,} a lowering in energy consumption for mass-transporting the suspension by shear. As shown previously, this is not only dictated by the viscosity reduction but also the flow path geometry \cite{acharya2023optimum}. In general, the viscosity reduction needs to be substantial enough so to compensate for the longer path one needs to take by doing a tacking motion. For frictionless grain suspensions the lowering in viscosity is considerable smaller than for frictional grains, mainly due to the constant shear-jamming packing fraction,  precluding any gain in dissipation per strain (see Fig. \ref{fig_diss} in the supplemental material \cite{supplemental}). 

{\it Microstructure:} Next we have a detailed look at the microstructure, examining how the number of contacts per particle at various tacking angles and packing fraction. In Fig. \ref{fig_nn}(a) and (b), we depict the average contact number as the packing fraction increases for different ASR angles for both frictionless and frictional particles. Notably, at each angle, this metric is higher with greater packing fractions, which is expected as a result of the increased likelihood of overlap as packing density increases. The observed decrease in the number of contacts with an increasing shear rotation angle aligns with the trends seen in Figs.~\ref{fig_jamming}(a) and (d), where a wider angle leads to the destruction of more contacts. Despite maintaining an exponential relationship akin to previous findings cited in Ref.~\cite{lin2016tunable}, the scaling factor $\eta/\eta_f \sim 14\exp(z)$ is not universally applicable, primarily due to variations in the contribution of hydrodynamics across different packing fractions. It is important to note that the total contribution includes hydrodynamics, which differs across packing fractions, necessitating its exclusion, especially when to map with the number of contacts. To address this, we plot the contact contribution to relative viscosity against the average coordination number for different packing fractions in Fig. \ref{fig_nn}(c). Strikingly, all curves for $\mu=0$ and $\mu=1$ converge separately into two different branches once we rescaled the contact contribution by a factor $f$, which depends upon $\theta$. We plot the $\theta$ dependency of this factor in the inset of Fig. \ref{fig_nn}(c). We fit these two branches keeping in mind two important facts, firstly, the stress contribution should diverge at the isostatic point which is $z=2d=6$ for frictionless and $z=d+1=4$ for infinite friction, secondly, the stress contribution from contact should be null at $z=0$. The frictionless branch fitted by $\eta/\eta_f\sim 25(\frac{z}{6-z})^2f(\theta)$ is rather simple, as $z_{iso}=6$ is standard. However, in the case of frictional as we have a friction coefficient $\mu=1$, which is far from infinite friction the isostatic point is not exact at $z_{iso}=4$ rather a slightly bigger value. Since the exact $z_{iso}$ is not easy to extract and will be slightly above $z=4$, we fit this with $\eta/\eta_f \sim 45(\frac{z}{4-z})^{2.5}f(\theta)$ and find that the fitted curve of these branches also shows a nice scaling including the deviation after a certain contact number.

{\it Discussion:}
We have shown that the shear-jamming point varies with cruising angle for frictional particles but not for frictionless.
We have found that for frictionless particles there exists distinct packing fraction (of around 59\%) where below the viscosity increases, \emph{i.e.}~thickens, upon increasing angle, while above the opposite happens, \emph{i.e.,}~thins. This effect is explained by the contact and hydrodynamics stresses opposing responses to an increasing angle. We have, furthermore, detailed how the viscosity and its various contributions, the number of contacts and contact network, varies after a shear rotation. 
This study forms a rigorous basis to study other shear protocols, with tailored mechanical responses and transport properties. 
Another way to tune rheology is to use orthogonal shear perturbations (OSP). In fact, these OSP are  the same as tacking \cite{acharya2023optimum}, where faster OSP shear rates correspond to wider angles. In this study, we have quantified how shear perturbations, including both alternating shear rotations and OSPs, affect the shear jamming point.
As shown in Fig.~\ref{parameter}(a), this impacts only the shear jamming point for frictional particles, approaching the frictionless point (close to the RCP) as the angle widens. 
Surprisingly, this shows many similarities to what has previously been done for perpendicular shear perturbations \cite{dong2020transition}.
Nevertheless, even if OSPs can alter a frictional rheology to a frictionless one, the number of contacts per particle still follows two different master curves, with a new hidden angular dependence $f(\theta)$, illustrating that there is still a fundamental difference between frictional or frictionless flows. \newline
Our work is important for many flows that involve a change in flow geometry, \emph{e.g.,} extrusion or landslides on varying slopes, and paves the way for more energy-efficient transport protocols exploiting geometry. 
\newline

{\it Acknowledgments:} MT is funded by the Swedish Research Council. PA acknowledges a generous scholarship from the Wenner-Gren Foundation.

\bibliographystyle{apsrev4-1}
\bibliography{vis}

\renewcommand{\thefigure}{S\arabic{figure}}
\newcommand{\beginsupplement}{%
        \setcounter{table}{0}
        \renewcommand{\thetable}{S\arabic{table}}%
        \setcounter{figure}{0}
        \renewcommand{\thefigure}{S\arabic{figure}}%
         \setcounter{equation}{0}
        \renewcommand{\theequation}{S\arabic{equation}}%
     }

\section*{\large Supplemental Material }

\subsection{Simulation details}
We simulate a bi-disperse $50:50$ mixture of hard particles with a diameter ratio of $1:1.4$ at fixed packing fractions $\phi$. The particles are immersed in iso-dense fluids, described as a continuum, with the viscosity $\eta_f$. The solid particles are modeled as athermal, non-inertial and spherical. The suspensions is confined in a three-dimensional periodic box using Lees-Edwards boundary conditions.

A particle undergoes three types of force and torque: Stokes drag, contact, and lubrication forces. The Stokes drag force and torque are given by,
\begin{eqnarray}
\nonumber
 \vec{F}_i^s &=& -6\pi\eta_f a_i (\vec{v}_i-\vec{v}_f),\\
\vec{\tau}^s_i &=& -8\pi \eta_f a_i^3 (\vec{\omega}_i - \vec{\omega}_f).
\end{eqnarray}
 Here, $a_i$, $v_i$, and $\omega_i$ are the radius, linear velocity, and angular velocity of particle $i$, while $v_f$ and $\omega_f$ are the fluid's linear and angular velocity (vorticity) and $\eta_f$ is the viscosity of the fluid. We chose parameters such that the Stokes number $\rho \dot \gamma \langle a\rangle^2/\eta_f\ll1$, where $\langle a\rangle$ denotes the average radius and $\rho$ the density of both solid particles and the fluid. We apply linear shear in various directions; therefore we generally have $\vec{v}_{f,\zeta}=\dot \gamma z \hat{\zeta}$, where $\hat{\zeta}$ is the unit vector in the flow direction and $z$ the gradient direction coordinate. In a corresponding way we have $\vec{\omega}_{f,\zeta}=\dot \gamma (\hat{\zeta} \times \hat{z}) /2$, $\hat{z}$ being the unit vector along the gradient direction.
 The contact force between particles is described by harmonic springs
 \begin{equation}
\vec{F}_{ij}=k_n\delta^{ij}_n \hat{n}^{ij}+k_t\delta^{ij}_t \hat{t}^{ij}.
\end{equation}        
Here $\delta^{ij}_n$ is the normal overlap between two particles while $\delta^{ij}_t$ is the relative tangential displacement. The springs' normal and tangential stiffness constants are denoted by $k_n$ and $k_t$.
We choose $k_t=(2/7)k_n$, and $k_n$ and $k_t$ values large enough such that our particles full-fill near-hard sphere conditions (typically $k_n/P \sim 10^4$).
The unit vector along the normal and the tangential direction of two particles are denoted by $\hat{n}^{ij}$ and $\hat{t}^{ij}$. The tangential forces for each contact are restricted by the Coulomb friction criteria $|\vec{F}^t_{ij}|<|\mu\vec{F}^n_{ij}|$, where, $\mu$ is the coefficient of friction. When a lubricating film separates two particles, the hydrodynamic interactions among particles rely on the resistance matrix formalism outlined in references \cite{mari2014shear,kim2013microhydrodynamics} and the code can be found in \cite{Ranga18}.
\subsection{Stress tensor calculation in new frame}
The stress tensor in a new frame can be obtained from the old one by performing a rotation of the stress tensor of the old frame as 
 $\sigma'= R^T \sigma R$, where $R$ is the rotation matrix about the gradient direction. Without loss of generality, we denote the flow, vorticity, and gradient directions in one of the frames by $x$, $y$ and $z$, respectively. With the rotation matrix the new stress component (in the flow/gradient plane) can be written as $\sigma_{fg}'=\sigma_{xz}\cos\theta +\sigma_{yz}\sin\theta$, where $fg$ stands for flow-gradient. Relative viscosities can then easily be obtained by dividing with $\eta_f \dot \gamma$, where $\eta_f$ is the fluid viscosity and $\dot \gamma$ the shear-rate. The later is kept constant and independent of the angle.

\begin{figure}[t!]
\includegraphics[width=1.0\linewidth]{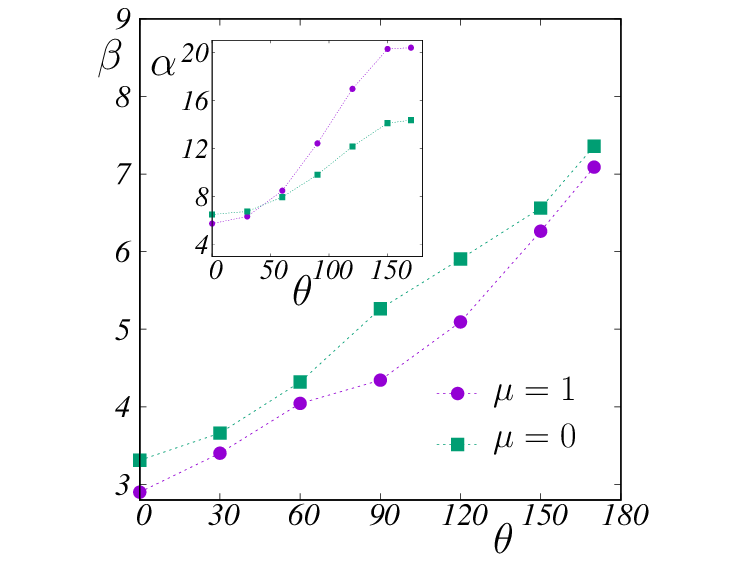}
\caption{Change in the exponent from the fitting of hydrodynamic contribution for both frictional and frictionless particles. (Inset) The prefactor from the fitting of hydrodynamic contribution for both frictional and frictionless particles. Both prefactor and exponents shows similar trend for both frictional and frictionless case. }
\label{fig_h_para}
\end{figure}

\begin{figure}[t!]
\includegraphics[width=1.0\linewidth]{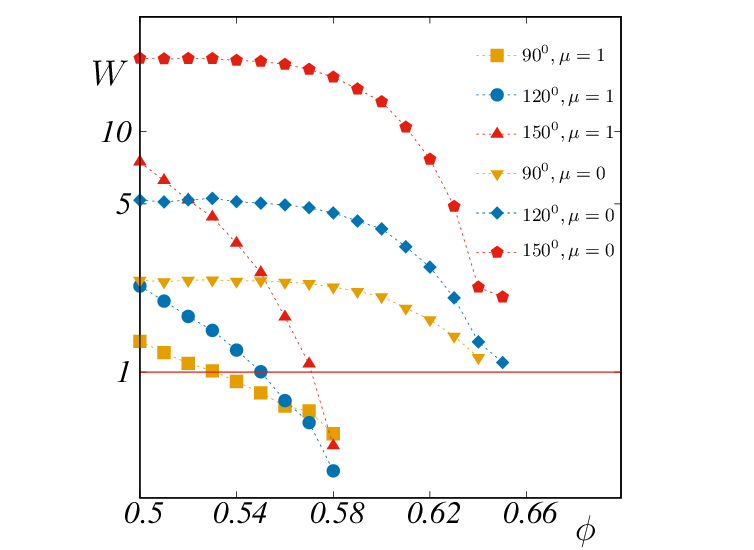}
\caption{Dissipation per unit strain for three different angle of shearing for both frictional and frictionless systems. We use $W = \frac{2}{(1+\cos\theta)} \frac{\eta}{\eta_{\rm ss}}$, from \cite{acharya2023optimum}. In frictionless systems, the value does not drop below 1, preventing any increase in dissipation per strain. In contrast, frictional systems exhibit a substantial gain after reaching a certain packing fraction for all three angles.}
\label{fig_diss}
\end{figure}


\begin{figure*}[t!]
\centering
\includegraphics[width=0.8\linewidth]{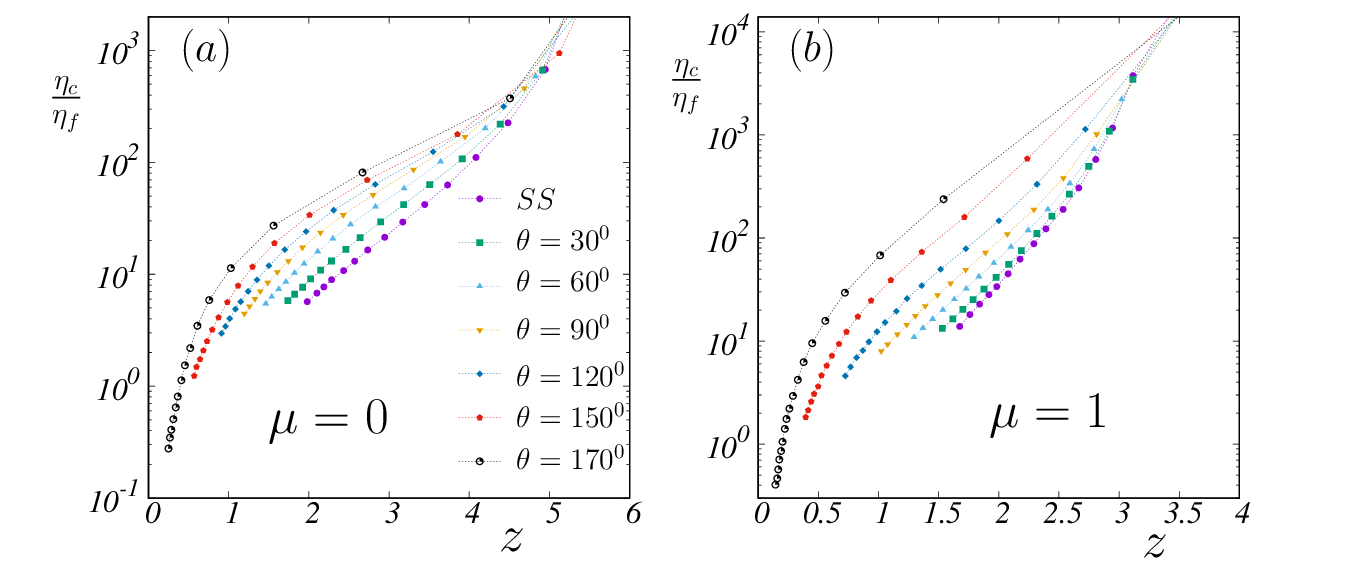}
\caption{ Contact contributions to the viscosity with contact number at various shear rotation angles for (a) $\mu=0$, (b) $\mu=1$.
}
\label{fig_zvseta}
\end{figure*}

\begin{figure*}[ht!]
\includegraphics[width=0.8\linewidth]{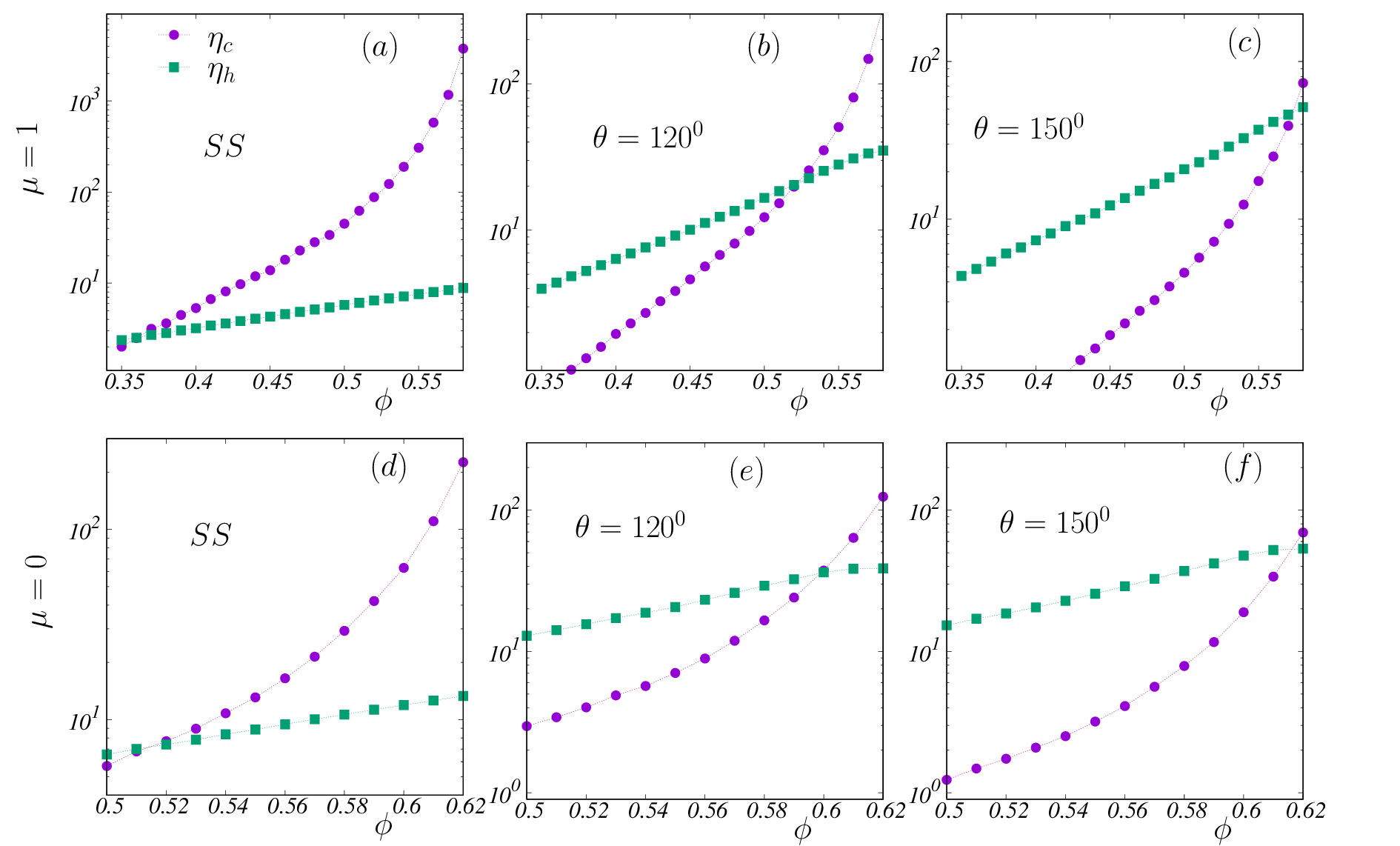}
\caption{The contact and hydrodynamic contributions as a function of packing fraction for various shear rotation angles (a,d) $\theta=0$, (b,e) $\theta=120^\circ$ and (c,f)  $\theta=150^\circ$. Top panel (a-c) corresponds to $\mu=1$ and bottom (d-f) to $\mu=0$.}
\label{fig_cross_SI}
\end{figure*}
\subsection{Parameters for the hydrodynamic fits}
In Fig.~\ref{fig_h_para}, we present the variation of the exponent obtained from the fitting of the hydrodynamic contribution for both frictional and frictionless particles. Additionally, in the inset we show the prefactor from the same fitting. Using the fitting form $\eta_h/\eta_f \sim \alpha (\phi/\phi_m)^\beta$, the simultaneous increase in both the prefactor and the exponent suggests an overall enhancement of the hydrodynamic contribution with increasing tacking angle, supporting the trends observed in Fig.~1(c) and (f) of the main manuscript. The crossing of the prefactor in the inset is due to our chosen $\phi_m=0.5$

\subsection{Dissipation}

In Fig.~\ref{fig_diss}, we plot the dissipation per strain for both frictionless and frictional cases at three different tacking angles. Once the viscosity is known, the dissipation per strain is calculated as $W = \frac{2}{(1+\cos\theta)} \frac{\eta}{\eta_{\rm ss}}$
following \cite{acharya2023optimum}. Since $W$ is defined as $W_{ASR}/W_{SS}$ a value less than 1 indicates a reduction in dissipation, signifying an energy gain in transporting the suspension over a steady state protocol. Notably, no such gain is observed for frictionless particles, which supports the trend shown in Fig. 1(b) of the main manuscript. In this case, all curves approach the same jamming point as in the steady state, preventing any substantial gain, unlike in the frictional case.

\subsection{Contact viscosity versus contact number at various angles}
In Fig.~\ref{fig_zvseta}(a) and (b), we plot the contact contribution to the relative viscosity alongside the coordination number for different rotation angles for frictionless and frictional particles, respectively. When applying the scaling factor discussed in the main text, all the curves in Fig. \ref{fig_zvseta}(a) and (b) collapse separately, forming two distinct branches, as shown in Fig.~3(c) of the main manuscript.

\subsection{Contact and hydrodynamic contributions relative to each other as a function of $\phi$ at various $\theta$'s}
In Fig.~\ref{fig_cross_SI}, we present the contact and hydrodynamic contributions to viscosity for three different angles, considering both frictional and frictionless particles. In all cases, a crossing point is observed, which shifts to higher values as the tracking angle increases. These crossing points correspond to those shown in Fig.~2(b) of the main manuscript.

\end{document}